# Privacy Issues in Voice Assistant Ecosystems

Georgios Germanos*, Dimitris Kavallieros*†, Nicholas Kolokotronis*, and Nikolaos Georgiou†

*University of Peloponnese, 22131 Tripolis, Greece
Email: {germanos, d.kavallieros, nkolok}@uop.gr
†Center for Security Studies (KEMEA), 10177 Athens, Greece
Email: {d.kavallieros, n.georgiou}@kemea-research.gr

*Abstract*—Voice assistants have become quite popular lately while in parallel they are an important part of smart-home systems. Through their voice assistants, users can perform various tasks, control other devices and enjoy third party services. The assistants are part of a wider "ecosystem". Their function relies on the users' voice commands, received through original voice assistant devices or companion applications for smartphones and tablets, which are then sent through the internet to the vendor's cloud services and are translated into commands. These commands are then transferred to other applications and services. As this huge volume of data, and mainly personal data of the user, moves around the voice assistant ecosystem, there are several places where personal data is temporarily or permanently stored and thus it is easy for a cyber attacker to tamper with this data, bringing forward major privacy issues. In our work we present the types and location of such personal data artifacts within the ecosystems of three popular voice assistants, after having set up our own testbed, and using IoT forensic procedures. Our privacy evaluation includes the companion apps of the assistants, as we also compare the permissions they require before their installation on an Android device.

## I. INTRODUCTION

As the Internet of Things is becoming more and more prevalent in our day-to-day life, there is a significant increase in the use of services and devices, the so-called *voice assistants*, also known as smart speakers or intelligent virtual assistants. In January 2019, more than 66 million smart voice assistants were installed in houses across the United States. The adoption rate in the country was at about 25%, and by 2022, it's expected to reach 55% [1].

Practically, a smart speaker is a type of wireless speaker, that comes with an integrated voice-activated digital assistant. Voice assistants perform tasks on behalf of a user: they make phone calls, answer general or specific questions (e.g. "what time does the next flight to New York depart?"), customize a user's daily schedule, set reminders, make product purchases on behalf of the user and even control IoT enabled devices (e.g. lights and locks) [2].

Other devices, mainly mobile phones and personal computers, may also function as voice assistants, through respective applications.

To achieve all the above mentioned tasks, it is necessary for them to become part of a wider ecosystem, in order to interact with the provider's cloud services and various other devices and applications. During their interaction, voice assistants exchange a large amount of data. Much of this data is considered personal information related to the user(s) of the ecosystem, which means that privacy issues arise.

In this paper, we focus on privacy issues, by examining the types and location of data and personal information that reside within a voice assistant ecosystem, while in parallel presenting existing IoT forensic techniques - which can also be used by a security professional or an investigator to have access to valuable evidence related to cyber attacks. Additionally, we compare the permissions required by the companion applications, before their installation on an Android mobile device, to verify if they are indeed necessary for the proper functioning of the voice assistant.

To verify the findings of our research, we built a testbed where we simulated the functioning of three very well-known assistants, Amazon's Alexa, Google's Assistant and Microsoft's Cortana.

The rest of the paper is organized as follows: In Section II we provide the background on voice assistant ecosystems, legal and privacy issues as well as Internet of Things forensic methodologies. Section III formalizes the methodology we followed to setup our testbed and the processes for the extraction and analysis of data. Section IV presents in detail our findings and an analysis on them, while Section V summarizes our work and outlines future work directions.

## II. BACKGROUND AND RELATED WORK

In this section we firstly describe a typical voice assistant ecosystem. Secondly, we discuss legal and privacy issues related to voice assistants. Last, we present methodologies that have been suggested for Internet of Things forensics procedures and techniques.

### A. Voice assistants

Voice assistants can be integrated into devices specifically designed for this purpose or work as an application on a smart phone, tablet or computer. A typical example of a widespread voice assistant is Amazon's Alexa. Alexa is the digital assistant

This project has received funding from the European Union's Horizon 2020 research and innovation programme under grant agreement no. 786698. The work reflects only the authors' view and the Agency is not responsible for any use that may be made of the information it contains.

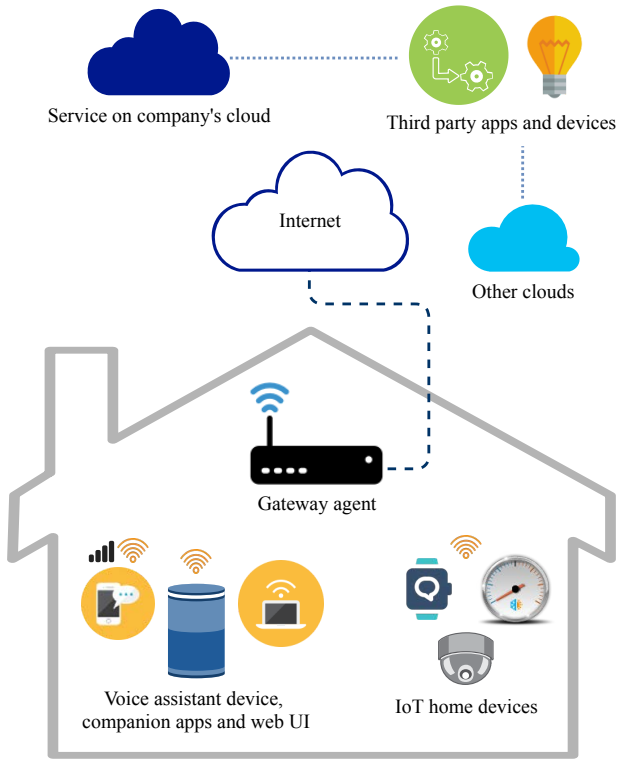

Fig. 1: A typical ecosystem of a voice assistant

that comes with the devices such as Echo, Echo Dot, Echo Show [3]. Other well-known voice assistants are Google's Assistant, and Microsoft's Cortana.

A typical ecosystem of a voice assistant is illustrated in Fig. 1. The user interacts directly, using his voice, with the specifically designed for this reason device of the company (e.g. Amazon Echo, Google Home, etc.), or through the companion app, which he has installed on one of his devices (mainly smart phone and tablet, but also laptop or desktop). His request is sent, through a network, to the company's cloud service, where further processing takes place. Then, the user's commands reach other devices and applications. The user finally receives the answer to his request through a voice message. In this way, the user can, for example, turn the lights of his house on, listen to his favorite music or order a taxi.

More in depth, the user interacts directly with the system using his voice. The device is enabled after it listens to the defined "wake up" word. It is designed to continuously listen to the sounds of the environment and start recording the user's voice (and command) only after the user speaks the "wake up" word. The file with the user's recorded command is sent, through the internet, to the cloud service of the provider, where extensive voice processing takes place, the voice is transformed into text and commands are identified. Depending on the commands, the provider either makes use of its own services to execute the task, or forwards it to other providers. This means that other smart IoT devices may receive commands from the provider. Then, back in the cloud, the provider prepares the response - result of its tasks - to the user, which is finally delivered to the user through the voice assistant, again in the form of audio.

This practically means that the user's data move around and may be located in the voice assistant device (hardware), the companion applications (client), the company's cloud (cloud) and the networks used for communication (network) [4].

### B. Legal and privacy issues

Within the European Union, the *General Data Protection Regulation* entered into force on May 25th 2018 [5]. Its purpose is to protect the data of the citizens of EU countries that are processed by individual or companies anywhere in the world, as well as any personal data that are processed by individuals or companies located within the territory of the EU. Among others, one of the priorities of GDPR is ensuring the respect and protection of *privacy*.

According to the GDPR, personal data means any information relating to an identified or identifiable natural person. Voice is also considered personal data and specifically biometric data, since it is related to the physical characteristics of a natural person, and can be used to identify individual. Processing means any operation or set of operations which is performed on personal data or on sets of personal data, such as collection, recording, organisation, structuring, storage, adaptation or alteration, retrieval, consultation, use and more.

It is clear that voice assistants do exactly the above mentioned processes, which means that, at least for voice assistants used by EU citizens or for processes of data that take place within the EU, the GDPR rules have to be respected. There are seven key principles set by the GDPR:

- Lawfulness, fairness and transparency
- Purpose limitation
- Data minimization
- Accuracy
- Storage limitation
- Integrity and confidentiality (security)
- Accountability

Article 5 of the GDPR requires that processing will be performed in a manner that ensures appropriate security of the personal data, including protection against unauthorised or unlawful processing and against accidental loss, destruction or damage, using appropriate technical or organisational measures.

An important principle of GDPR is "privacy by design", as described in article 25. The controller must, both at the time of the determination of the means for processing and at the time of the processing itself, implement appropriate technical and organisational measures, which are designed to implement data-protection principles, in an effective manner. In parallel, the controller should integrate the necessary safeguards into the processing in order to meet the requirements of the GDPR and protect the rights of data subjects. For the controller to implement the appropriate technical and organizational measures, he should take into account the state of the art, the cost of implementation and the nature, scope, context and purposes of processing as well as the risks of varying likelihood and severity for rights and freedoms of natural persons posed by the processing.

*C. IoT forensics*

The issue of digital forensics in voice assistants has attracted the attention of many researchers as it is a major challenge. As a rule, scientific research focuses on studying traditional devices, such as hard drives or cell phones. To date, there is no serious need for digital forensics of voice search devices. However, it is almost certain that these devices, as part of the Internet of Things, will gain the attraction either Law Enforcement Authorities or specialists in the field of digital forensics. The digital examiner will be asked to study the function of his voice assistant, his memory, his connections and more. Due to the specificity of each voice assistant device, each one requires a specialized approach [6], [7].

To date, there is no software or tool specifically tailored to conduct forensic investigations on voice assistants. Each case is considered individually, however some IoT device testing models have been suggested that could be applied to voice assistants as well. It is certain that the incorporation of the 'forensics by design' principle during the design phase of a voice assistant would greatly solve the research question on these devices.

There are a few methodologies that have been proposed for IoT forensics procedures and techniques. Zawoad et al. [8] proposed a Forensics-aware IoT (FAIoT) model for supporting reliable forensics investigations in the IoT environment. They identified IoT forensics as a combination of three digital forensics schemes, on the basis of the IoT ecosystem: device level forensics, network forensics, and cloud forensics. Their model includes a centralized trusted evidence repository to facilitate the process of evidence collection and analysis. At the same time, reliability of the evidence is also achieved, thanks to the application of the secure logging scheme.

Perumal et al. [9] proposed a novel model, which included authorization, planning, chain of custody, analysis and storage, and was based on triage model and 1-2-3 zone model for volatile based data preservation.

In [10], Kebande et al. proposed a generic Digital Forensic Investigation Framework for IoT (DFIF-IoT), which was expected to support future IoT investigative capabilities. The proposed framework would comply with the ISO/IEC 27043:2015 which is an international standard for information technology, security techniques, incident investigation principles, and processes.

A procedure for digital evidence acquisition model for IoT forensic was provided as a theoretical framework by Harbawi et al. in [11]. One aspect of the procedure was the deployment of Last-on-Scene algorithm that would improve traceability and reduce the overhead as well as digital forensic analysis complications. Then, a revision was made for IoT forensic in order to ensure the usability of the proposed enhanced acquisition procedure. Last, a management platform concept was also proposed.

Meffert et al. presented a primary account for a general framework and practical approach, called "Forensic State Acquisition from Internet of Things (FSAIoT)" [12]. The method provided a general solution towards device state acquisition. It would not physically acquire the memory of the devices themselves. FSAIoT consisted of a centralized Forensic State Acquisition Controller (FSAC) employed in three state collection modes: controller to IoT device, controller to cloud, and controller to controller.

In their work [13], Dorai et al. examined the forensic artifacts produced by Nest devices, and specifically the logical backup structure of an iPhone used to control these devices. For this reason, they built an open-source forensic tool, the "Forensic Evidence Acquisition and Analysis System (FEAAS)", that would consolidate evidentiary data into a readable report that could infer user events.

Goudbeek et al. proposed a forensic investigation framework for the smart home environment, which could also be used as a quick reference guide for digital forensic investigators working on future home automation systems. The framework included seven phases, during which a range of artefacts of forensic relevance could be recovered to inform forensic investigations [14]. The same framework has been used by Tristan et al. to perform forensic investigation in smart speakers [15].

III. METHODOLOGY

In order to examine the privacy issue more in depth, we set up our own testbed for three different voice assistants. Furthermore, a forensic analysis took place in order to identify any information related to the identity or the activities of the user.

*A. Testbed setup*

As a means to have a better understanding of the three major voice assistants ecosystems, we created our own testbed, in a home environment, which we used for a period of almost 8 weeks, from April 1st 2020 to May 31st 2020. Our testbed is presented in Fig. 2.

More specifically, we used a Xiaomi Redmi Note 6 Pro mobile phone, with Android 9 installed on it, a Lenovo Yoga 730 laptop, with Windows 10 installed on it, as well as a Raspberry Pi 4, with Raspberry Pi OS (ex. Raspbian) installed on it, to simulate the functioning of voice assistant devices. These devices had wireless internet access through a typical home router device (ZTE H108N). The mobile phone had also access to the internet through 4G connection.

We installed the companion mobile apps "Amazon Alexa", "Google Assistant – Get Things Done, Hands-free", as well as "Microsoft Cortana – Digital Assistant" that were available on Play Store. We installed the applications for Amazon's Alexa and Microsoft's Cortana on the laptop. We also installed Amazon's Alexa and Google's Assistant on the Raspberry Pi device [16], [17]. For the analysis of the storage of the Raspberry Pi, we primarily used the FTK Imager while, for the clients, methods for mobile and computer forensics were used.

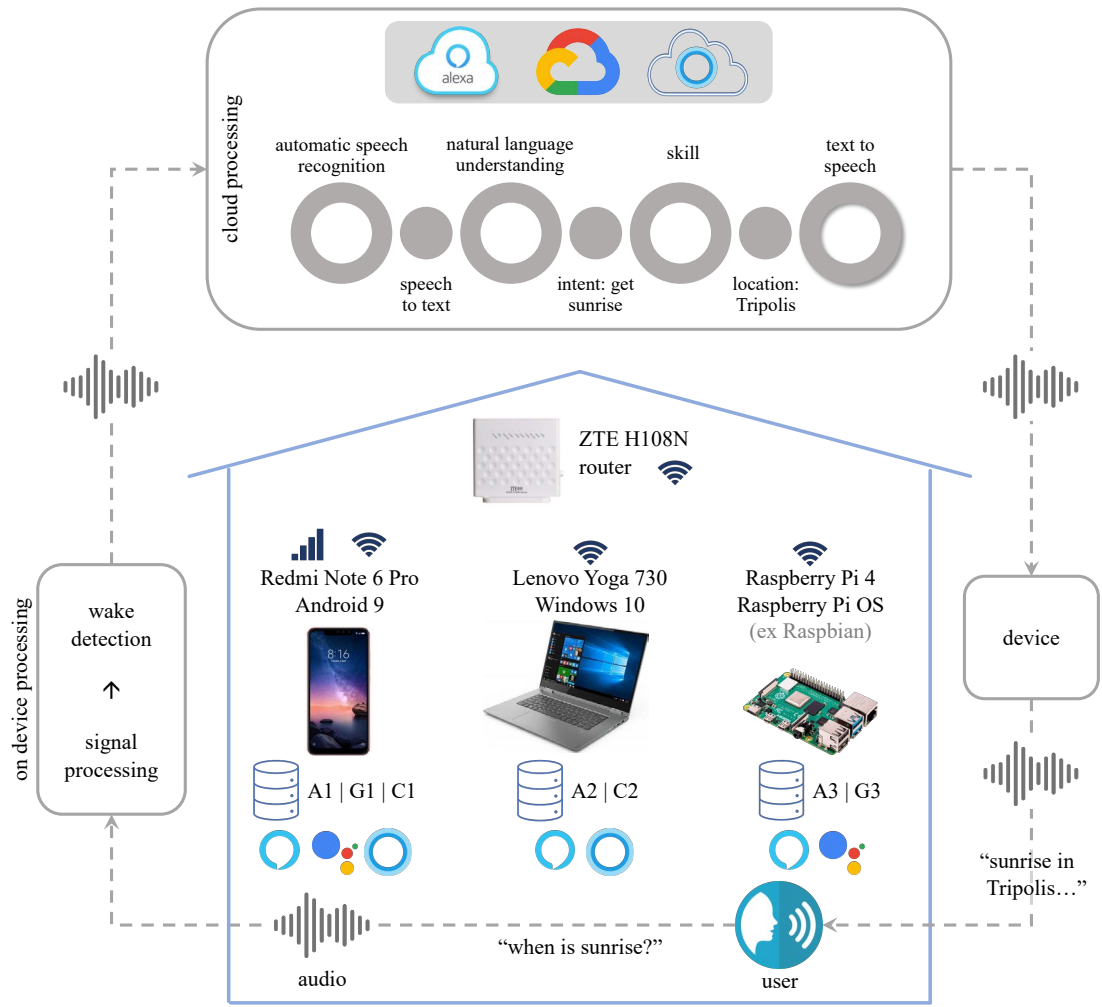

Fig. 2: The testbed used for the evaluation of the voice assistants ecosystems' privacy

To better understand the location of the artifacts of Table I and II, we use A for Alexa, G for Assistant and C for Cortana and 1 for the mobile device, 2 for the laptop and 3 for the Raspberry Pi. A1 in Fig. 2 refers to the artifacts related to Alexa, found on the mobile device, A2 refers to the artifacts related to Alexa, found on the laptop, etc.

### B. Testing process

With a view to accurately test the proper functioning of the ecosystems, we created three different accounts (profiles), one for each ecosystem, through their respective web interface.

In practice, we interact, using our voice, with each ecosystem through the mobile phone, the laptop and the Raspberry Pi (end device): each of our oral questions – requests travels through the end device to the internet and then to the cloud of each company, where further processing takes place. We then receive, in voice format, the answer to our request through the end device.

The testing included various commands such as, creation of To-do list, open applications, listen to music, add items to the calendar, identify traffic conditions between two locations, create shopping list, open images and videos stored in the end device, etc.

At the end of the testing, we performed information gathering and analysis in order to identify possible artifacts correlated with the user's privacy. The main activities employed for the identification of information stored in the devices of our testbed were, (a) image acquisition and analysis of the Raspberry Pi storage, (b) manual search for the file path and the location of the folders the voice assistants store data (in the laptop and android), (c) usage of APIs for information retrieval from the respective cloud and (c) literature review.

The image of the Raspberry Pi storage was acquired using the FTK Imager, while for the analysis the Forensic Toolkit (FTK) was used.

It is important to highlight that, in order to retrieve data through the APIs, we assumed that we already had access to the user's account.

TABLE I: Location of voice assistant artifacts in our testbed

| Location | A: Alexa (Amazon) | G: Assistant (Google) | C: Cortana (Microsoft) |
|---|---|---|---|
| User registration | alexa.amazon.com | assistant.google.com | www.microsoft.com/en-us/cortana |
| 0: Cloud | https://www.amazon.com/hz/mycd/digital-console/privacysettings | https://myactivity.google.com/item | https://account.microsoft.com/privacy/activity-history |
| 1: Mobile | /Android/data/com.amazon.dee.app | /Android/data/com.google.android.apps.chromecast.app | /Android/data/com.microsoft.cortana/files |
| 2: Laptop | \\$\langle$User$\rangle$\\AppData\\Local\\Packages\\57540AMZNMobileLLC.AmazonAlexa_22t9g3sebte08 | | \\$\langle$User$\rangle$\\AppData\\Local\\Packages\\Microsoft.Windows.Cortana_cw5n1h2txyewy |
| 3: Raspberry Pi | /root/home/pi/db/... | /root/.config/googlesamples-assistant/device_config.json | |

TABLE II: User's artifacts found in various ecosystems

| Artifact | A: Alexa (Amazon) | G: Assistant (Google) | C: Cortana (Microsoft) |
|---|---|---|---|
| Access information | | yes (password, phone number) | |
| Activities | yes (to-do list, shopping list) | yes (files on Google drive) | yes (photo albums, apps installed in devices) |
| Calendar details | yes (alarms, reminders) | yes | yes |
| Contacts | yes | yes | yes (incl. relationships if set up) |
| Emails | yes | yes | yes |
| History: activities | yes | yes (searches, viewed videos/ads) | yes (communications from messages, apps) |
| History: audio input | yes (voice commands) | | yes (voice commands) |
| History: browsing | yes | yes | yes |
| Interests | yes (restaurants, sports, traffic) | | yes (sports, news, etc. from apps, services) |
| Location data | yes | yes | yes (also linked to users' requests) |
| Multimedia | yes (photos via Amazon photos) | yes (photos, videos) | yes (music, books, podcasts) |
| Personal details | yes (name, birthday, gender) | yes (name, birthday, gender) | yes (name, nickname, age, family members) |
| Systems connected | yes (apps, devices, services) | yes (apps, devices, services) | yes (apps, devices, services in *home* app) |

## IV. FINDINGS AND ANALYSIS

In this section we present the findings of our experiments within the testbed and at the same time an analysis from a privacy perspective.

### A. Personal data and other information

In their official websites, Amazon, Google and Microsoft explain the types of data that they process [18], [19], [20].

We were able to locate and identify various types of artifacts, related to our activities. The location of the artifacts is presented in Table I.

The manual search on the A2 and C2 revealed the locations where Alexa and Cortana, respectively, store information in a laptop device. For example, in Cortana the search history of the user, applications that he had used and voice command history, as well as the respective audio files were identified [21], [22].

It is important to highlight that the file path presented in Table I regarding the mobile device can only be found if the mobile device is rooted. The mobile used for A1/G1/C1 was not rooted thus, no manual search could be conducted.

Using the unofficial API presented in [23], we could identify plethora of personal information, including locations, the email of the account and photos, among other artifacts. Regarding Alexa we found very limited amount of information.

As it was presented in the previous section, forensic analysis was conducted in A3 and G3. The main objective of the analysis was to identify the presence/absence of artifacts related to either Alexa or Assistant (Google). The following artifacts were identified regarding Alexa:

a) device serial number,
b) the client ID,
c) the product ID,
d) client ID name,
e) profile ID
f) name of the user
g) secret of the user

Specifically, artifacts (d)–(g) were identified in the cache memory. Fig. 3a and Fig. 3b are depicting this information. Regarding Google Assistant, we were able to identify the following artifacts:

a) client secret, and
b) client ID.

These finding are presented in Fig. 3c.

A summary of these artifacts is presented in Table II. More in depth, we were able to identify access information (e.g. password, phone number), various user activities (e.g. to-do list, shopping list, photo albums, apps installed on devices), details related to calendars (e.g. alarms, reminders), contacts, email accounts, history of user's activities (e.g. searches, viewed videos/ads, communications from messages, apps, browsing, audio inputs), interests (e.g. restaurants, sports, traffic, news), location data, multimedia files (photos, videos, music, books, podcasts), personal details (such as name, birthday, gender, nickname, age) and systems connected (apps, devices, services).

### B. Applications' permissions

In Table III we present the permissions that each voice assistant application (Alexa, Assistant, Cortana) requests, before

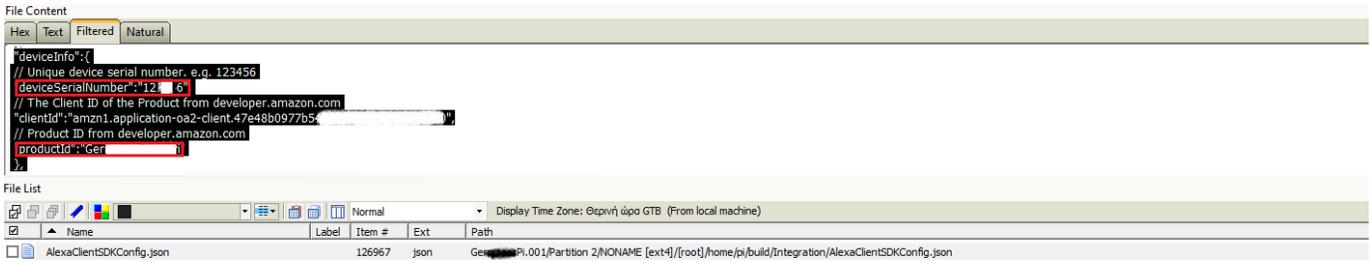
(a) The device serial number and product ID used by Amazon Alexa in the testbed

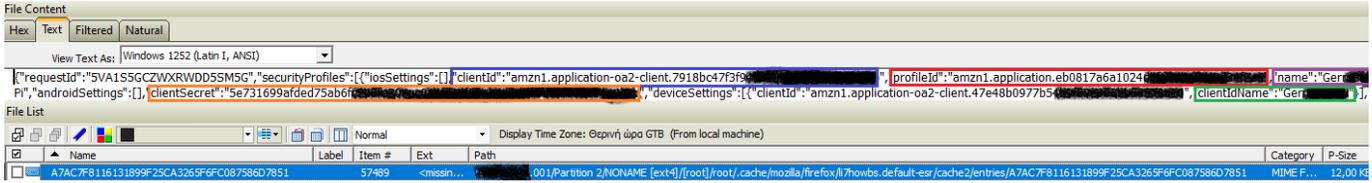
(b) The client ID, profile ID, user name, client secret, client name used by Amazon Alexa in the testbed

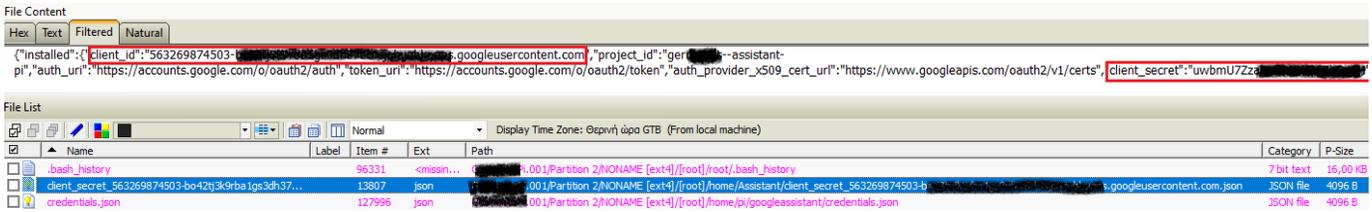
(c) The client ID and secret used by Google Assistant in the testbed

Fig. 3: Examples of artifacts being identified during the analysis of the Raspberry Pi images with the Forensic Toolkit (FTK)

its installation on Android devices.

By exploiting known or zero-day vulnerabilities of these features, someone could gain unauthorized access to the device and its data. The permissions required by the assistant applications are numerous.

Within the Android environment, permissions characterized as "dangerous" require user agreement. How Android asks for dangerous permissions depends on the application version and the system version targeted by the application [25].

The last column in Table III presents if the permission is among the dangerous ones (if the permission is dangerous, D is present).

The assistants need permissions to use various features of the mobile device, nevertheless, Cortana's list of permissions looks extensive. At the same time, it should be noted that Google's Assistant permissions look smaller in number, but this is due to the fact that there are usually many other Google's applications installed on Android devices [24]. This means that, most probably, Google already has the "missing" permissions of the Table.

In terms of statistics, Alexa asks for 17, Assistant (Google) asks for 6 and Cortana asks for 22 dangerous permissions. This does not necessarily mean that Assistant gives importance in user's privacy, as we already mentioned that many of the Assistant's permissions which are not present in the Table have already been provided to Google through other applications.

Table IV below presents the artifacts and information found in each assistant, on the different "devices" as depicted on Table I, namely, the cloud of each assistant, the Raspberry Pi installation – A3, G3 (only for Alexa and Assistant) and the respective clients – A1, A2, G1, G3, C1 and C3.

It also provides, in the case of Android clients, a correlation among the identified artifacts and the permissions depicted on Table III. This list is not an exhaustive as it is based on the setup and testing described in Section III.

## V. CONCLUSIONS AND FUTURE WORK

In our work we examined privacy issues in voice assistant ecosystems. After describing a typical voice assistant ecosystem, we presented the existing legislation on personal data processing within the EU and its privacy aspects. We then made a literature review of existing IoT forensics methodologies.

Our main goal was to identify the types and location of data that reside within a voice assistant ecosystem. For this reason, we built our own testbed where we simulated the functioning of Amazon's Alexa, Google's Assistant and Microsoft's Cortana voice assistants.

We also compared the permissions required by the voice assistant companion applications before their installation on an Android device and their relation to privacy.

In the future, our intention is to examine ways of enhancing privacy in voice assistant ecosystems, as well as further study attacks on voice assistants and formulate a novel forensic methodology for artifacts in voice assistant ecosystems.

TABLE III: The permissions requested by voice assistants' mobile application during their installation on an Android device and dangerous ones [25]. The notation 'A', 'G' and 'C' denotes Alexa (Amazon), Assistant (Google), and Cortana (Microsoft) respectively. Dangerous permissions are indicated by the letter 'D' in the last column.

| ID | Permission | A | G | C | | ID | Permission | A | G | C | |
|---|---|---|---|---|---|---|---|---|---|---|---|
| | **Identity** | | | | | | **SMS and MMS** | | | | |
| 1.1 | Read your own contact card | | | × | | 11.1 | Read text messages | × | | × | D |
| 1.2 | Find accounts on device | × | × | × | D | 11.2 | Receive text messages | × | | × | D |
| 1.3 | Create accounts and set passwords | × | | × | | 11.3 | Send text messages | × | | × | D |
| 1.4 | Add or remove accounts | × | × | × | | 11.4 | Edit text messages | | | × | |
| 1.5 | Use accounts on the device | × | | × | | | **Calendar** | | | | |
| | **Wi-Fi connection information** | | | | | 12.1 | Read events and private information | | | × | D |
| 2.1 | View Wi-Fi connections | × | × | × | | 12.2 | Add/modify calendar events | | | × | D |
| 2.2 | Connect and disconnect from Wi-Fi | × | × | × | | 12.3 | Send emails without users' knowledge | | | × | D |
| 2.3 | Allow Wi-Fi Multicast reception | × | | | | | **Device and app history** | | | | |
| | **Photos, media, files** | | | | | 13.1 | Retrieve running apps | | | × | |
| 3.1 | Modify/delete photos, media, files | × | | × | D | | **Other** | | | | |
| 3.2 | Read photos, media, files | × | | × | D | 14.1 | Receive data from Internet | × | × | × | |
| | **Microphone** | | | | | 14.2 | Run at startup | × | × | × | |
| 4.1 | Record audio | × | × | × | D | 14.3 | Full network access | × | × | × | |
| 4.2 | Change audio settings | × | × | × | | 14.4 | Control vibration | × | | × | |
| | **Device ID and call information** | | | | | 14.5 | Prevent device from sleeping | × | × | × | |
| 5.1 | Read phone status and identity | × | | × | D | 14.6 | Pair with Bluetooth devices | × | × | × | |
| | **Camera** | | | | | 14.7 | Change network connectivity | × | × | × | |
| 6.1 | Take pictures and videos | × | × | × | D | 14.8 | View network connections | × | × | × | |
| | **Contacts** | | | | | 14.9 | Access Bluetooth settings | × | × | × | |
| 7.1 | Find accounts on device | × | × | × | D | 14.10 | Read Google service configuration | | × | | |
| 7.2 | Read your contacts | × | | × | D | 14.11 | Hot word detection | | | × | |
| | **Phone** | | | | | 14.12 | Download files without notification | | | × | |
| 8.1 | Directly call phone | × | × | × | D | 14.13 | Full license to interact across users | | | × | D |
| 8.2 | Read phone status and identity | × | | × | D | 14.14 | Update component usage statistics | | | × | |
| 8.3 | Reroute outgoing calls | | | × | D | 14.15 | Read home settings and shortcuts | | | × | |
| | **Storage, media files** | | | | | 14.16 | Toggle sync on and off | | | × | |
| 9.1 | Modify/delete USB storage contents | × | | × | D | 14.17 | Install/uninstall shortcuts | | | × | |
| 9.2 | Read USB storage contents | × | | × | D | 14.18 | Read sync settings | | | × | |
| | **Location-based services** | | | | | 14.19 | Modify system settings | | | × | |
| 10.1 | Precise location (GPS, network-based) | × | × | × | D | 14.20 | Disable your screen lock | | | × | |
| 10.2 | Approximate location (network-based) | × | | × | D | 14.21 | Set an alarm | | | × | |
| 10.3 | Additional location provider commands | | | × | | 14.22 | Draw over other apps | | | × | |

Furthermore, part of our future work is to tamper with original assistant devices (e.g. use Alexa instead of Raspberry Pi), connected to multiple smart devices (e.g. smart lamps). Moreover, we intend to use rooted mobiles, mobile mirroring for un-rooted devices and emulator for different types of OS of mobiles. This approach will allow us to investigate for stored artifacts in different types of setups and compare them while at the same time we will be able to compare the results between the different testbeds.

TABLE IV: Correlation between the identified artifacts and the permissions depicted on Table III

| Artifact | A: Alexa (Amazon) | G: Assistant (Google) | C: Cortana (Microsoft) |
|---|---|---|---|
| Access information |  | 1.x |  |
| Activities | 3.x, 9.x | 3.x, 9.x | 3.x, 13.x, 14.19 |
| Calendar details | 12.x, 14.21 | 12.x | 12.x |
| Contacts | 7.x | 7.x | 7.x |
| Emails | 14.1 | 14.1 | 14.1 |
| History: activities | 3.x | 3.x, 9.x | 3.x, 11.x, 14.1 |
| History: audio input | 3.x, 4.x, 9.x |  | 3.x, 4.x, 9.x |
| History: browsing | 14.1 | 14.1 | 14.1 |
| Interests | 14.1 |  | 14.1 |
| Location data | 10.x, 14.1 | 10.x, 14.1 | 10.x, 14.1 |
| Multimedia | 3.x, 9.x | 3.x, 9.x | 3.x, 9.x |
| Personal details | 1.1 | 1.1 | 1.1–1.2, 7.x |
| Systems connected | 2.x, 13.x, 14.1, 14.6–14.8 | 2.x, 13.x, 14.1, 14.6–14.8 | 1.2, 2.x, 13.x, 14.1, 14.6–14.8, 14.15 |